\documentclass[]{spie}  
\usepackage[]{graphicx}
\usepackage{url}
\usepackage{amsmath, amssymb, bm}
\usepackage{aas_macros}
\usepackage{hyperref}
\usepackage{subcaption}

\title{Post-processing of the HST STIS coronagraphic observations} 

\author{Bin~Ren\supit{a}, Laurent~Pueyo\supit{b}, Marshall~D.~Perrin\supit{b}, John~H.~Debes\supit{b}, \'Elodie Choquet\supit{c*}
\skiplinehalf
\supit{a} Johns Hopkins University, 3400 North Charles Street, Baltimore MD 21218, USA; \\
\supit{b} Space Telescope Science Institute, 3700 San Martin Drive, Baltimore MD 21218, USA; \\
\supit{c} Jet Propulsion Laboratory, California Institute of Technology, 4800 Oak Grove Drive, Pasadena, CA 91109, USA\\
}

\authorinfo{* Hubble Fellow\\  \indent Send correspondence to B.R.: ren@jhu.edu}

\begin{document} 
\maketitle 

\begin{abstract}
In the past 20 years, the {\it Hubble Space Telescope} ({\it HST}) STIS coronagraphic instrument has observed more than 100 stars, obtaining more than 4,000 readouts since its installment on {\it HST} in 1997 and the numbers are still increasing. We reduce the whole STIS coronagraphic archive at the most commonly observed positions (Wedge A0.6 and A1.0) with new post-processing methods, and present our results here. We are able to recover all of the $32$ previously reported circumstellar disks, and obtain better contrast close to the star. For some of the disks, our results are limited by the over subtraction of the methods, and therefore the major regions of the disks can be recovered except the faintest regions. We also explain our efforts in the calibration of its new BAR5 occulting position, enabling STIS to explore inner regions as close as $0.2''$.
\end{abstract}



\section{INTRODUCTION}
\label{sec:intro}  

Planets are formed in protoplanetary disks through core-accretion or gravitational instability \cite{chambers10, dagenlo10}. Protoplanetary disks have typical lifetimes of a few times of $10^6$ yr \cite{wyatt08}, during which the protoplanetary disks evolve and the formation and migration of planets occur \cite{roberge10}. Observation and modeling of these disks help us better understand the structure, composition, and evolution of planetary systems. 

One of the most direct ways to study protoplanetary disks is analyzing their morphology, which can be obtained with high contrast imaging. An accurate recovery of disk morphology can help us in the understanding of disk properties in several ways. First, we can retrieve their surface brightness profiles, and reveal possible asymmetric structures and the traces of complex dynamical structures (spiral arms, jets, clumps, etc. See Refs.~\citenum{dong15} and \citenum{dong16} for some canonical examples). Second, we will be able to study the evolution of them on short timescales (see Ref.~\citenum{debes17a} for yearly evolution of the TW Hydrae disk). Third, the morphology of disks in simulation also indicates the possible existence of unseen planets that are perturbing the circumstellar disk structure and create observable signatures\cite{rodigas14, lee16, nesvold16, dong17}.

When the protoplanetary disk dissipates and planet formation completes, the mutual collision between the leftover planetesimals may lead to a second generation dust --- a debris disk\cite{wyatt08, roberge10}. The grains in debris disks have dynamical lifetimes much shorter than the ages of their host stars \cite{gillett86}, and they continue interacting with the stellar radiation, and the inter-stellar medium, as well as other planetary objects in the system\cite{lee16, nesvold16}. With the spectrum of a disk, we are able to obtain some of its compositions \cite{harker02, gail04, roberge06}; however, it cannot relate the compositions to spatial locations. For a debris disk, understanding its composition and grain size distribution may help us locate its origin \cite{stark14}, which can also be studied via the radiative transfer modeling of them (e.g., Ref.~\citenum{hung15}).

Detection and characterization of these circumstellar disks (i.e., protoplanetary and debris disks) with direct imaging rely on excellent design of telescope instruments and state-of-the-art post-processing of observations. Well-designed instruments are able to stabilize the temporally varying noise in telescope exposures, creating quasi-static features \cite{perrin04, hinkley07, perrin08, soummer07, traub10}. These quasi-static features (speckles), together with stellar point spread function (PSF), can be empirically modeled and removed by post-processing techniques, revealing circumstellar disks and exoplanets \cite{marois06, lafreniere07, lafreniere09, soummer12}.

Post-processing techniques start with the concept of differential imaging. When the stellar PSF and the speckles are stable and do not vary over time, they can be removed by subtracting another PSF image. As HST is relatively stable, observing strategies with HST included the observation of a reference star to enable PSF subtraction (Reference-star Differential Imaging, RDI). This classical RDI method have been extensively used, yielding multiple results especially for the brightest disks \cite{smith84, grady10, schneider14, schneider16, debes13, debes17}. However, for ground-based telescopes where the speckles do vary, several techniques have been proposed to resolve this (e.g., Angular Differential Imaging, ADI\cite{marois06}; Spectral Differential Imaging, SDI\cite{biller04}).

Current successful removal of both stellar PSF and speckles are based on the utilization of innovative statistical methods \cite{mawet12}. Using a large sample of references, there are two widely-known advanced post-processing algorithms: one is the Locally Optimized Combination of Images algorithm (LOCI)\cite{lafreniere07}, the other is the Karhunen-Lo\`eve Image Projection algorithm (KLIP)\cite{soummer12, amaya12}. The two methods are closely related, aiming at extracting maximum target signal and minimize noise using a reference library of covariant images\cite{savransky15}. Both of them have made a lot of new discoveries even with archival data \cite{lafreniere09, soummer12, soummer14, choquet14, choquet16, mazoyer14, mazoyer16}, however when the astrophysical signals overlap with the speckle noises and therefore cannot be separated out, both LOCI and KLIP will bias the astrometry, morphology and photometry of the exoplanets or circumstellar disks\cite{marois10, pueyo12}, and forward modeling has to be performed \cite{esposito14, wahhaj15, follette17, soummer12, choquet16, choquet17}. Due to its greater computational efficiency, as well as excellent controllability in forward modeling, we choose KLIP as our way of data reduction in this article.

{\it HST} has had many coronagraphs over the past 26 years, but currently the Space Telescope Imaging Spectrograph (STIS) instrument is the only one still operational. For the NICMOS coronagraph, the Archival Legacy Investigations of Circumstellar Environments (ALICE) project has systematically and consistently reprocessed its entire coronagraphic archive, and achieved much improved contrasts particularly at small separations\cite{soummer14, choquet14}. This led to the recovery of many new images of disks and new discoveries of point source candidate companions\cite{choquet16, choquet17}. The goal of our work presented in this article is to 1) extend this approach to the STIS archive, in order to see if reference library based PSF subtractions can achieve better performances and if so in what parts of parameter space; and 2) explore the limit of the recently identified BAR5 location for exoplanet imaging. Since this is the first large exploration of the whole STIS archive in this manner, we also have to investigate questions regarding how best to systematically and automatically align and process the data, and to assess how well the PSF subtraction performance works and what factors may set systematic limits. 

\section{Data Acquisition and Preparation}
\begin{figure}[htb!]
\begin{center}
\begin{tabular}{c}
\includegraphics[width=0.5\textwidth]{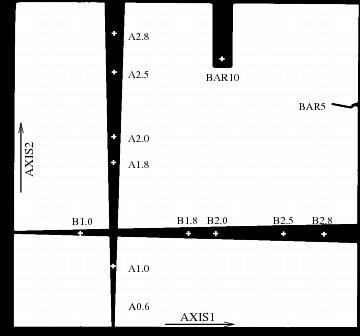}
\end{tabular}
\end{center}
\caption
{ \label{fig:stis-coron} Occulting positions for coronagraphic imaging with STIS\cite{stisIHB17}. The four major wedges are: A, B, BAR10, and BAR5. The numbers after the letters for A and B stand for the width across the wedge in arcseconds ($''$). The field of view for the STIS CCD is $52''\times52''$ ($1024\times1024$ pixel).}
\end{figure}

\begin{figure}[htb!]
\begin{center}
\begin{tabular}{c}
\includegraphics[width=0.45\textwidth]{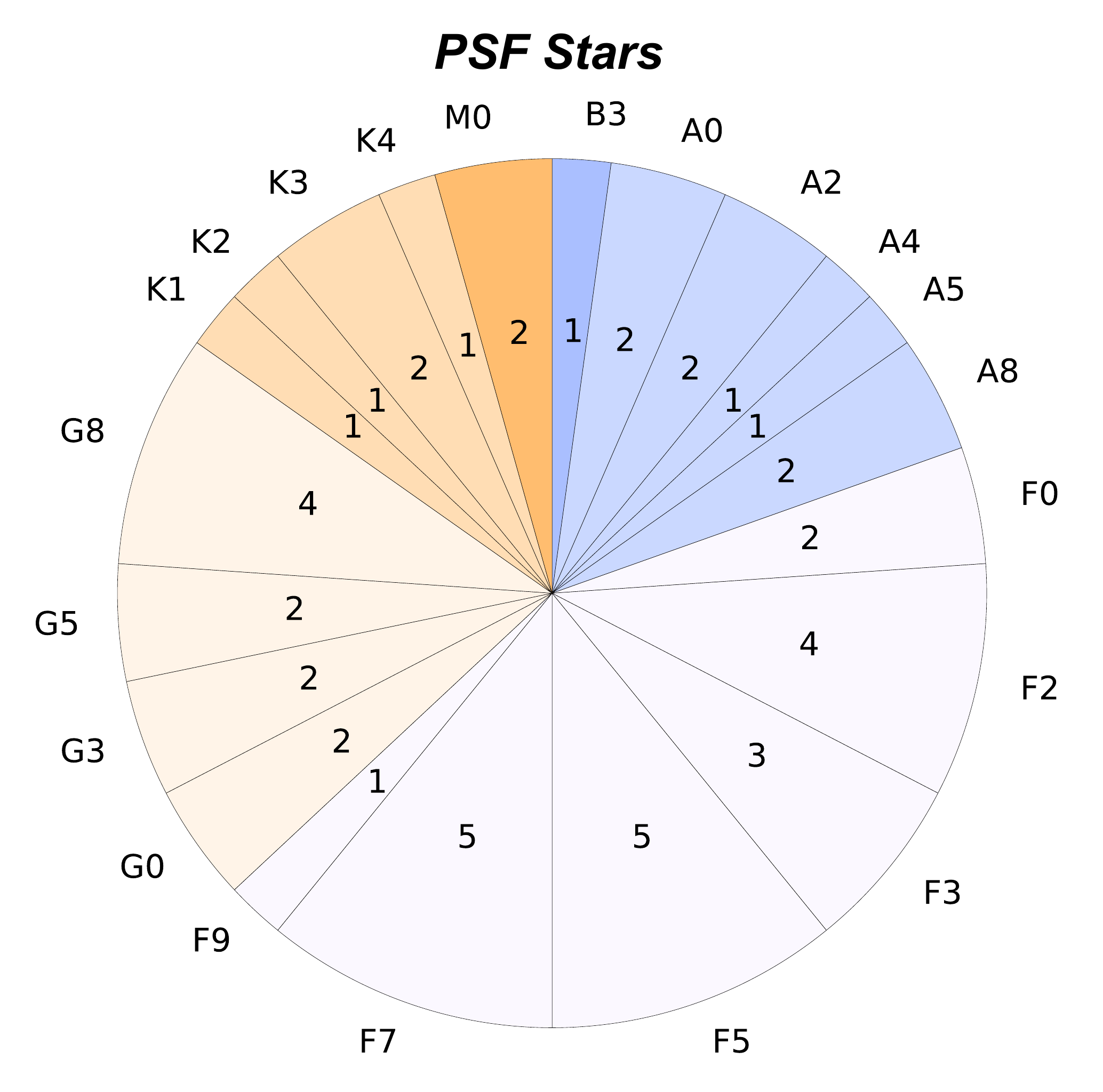}
\includegraphics[width=0.45\textwidth]{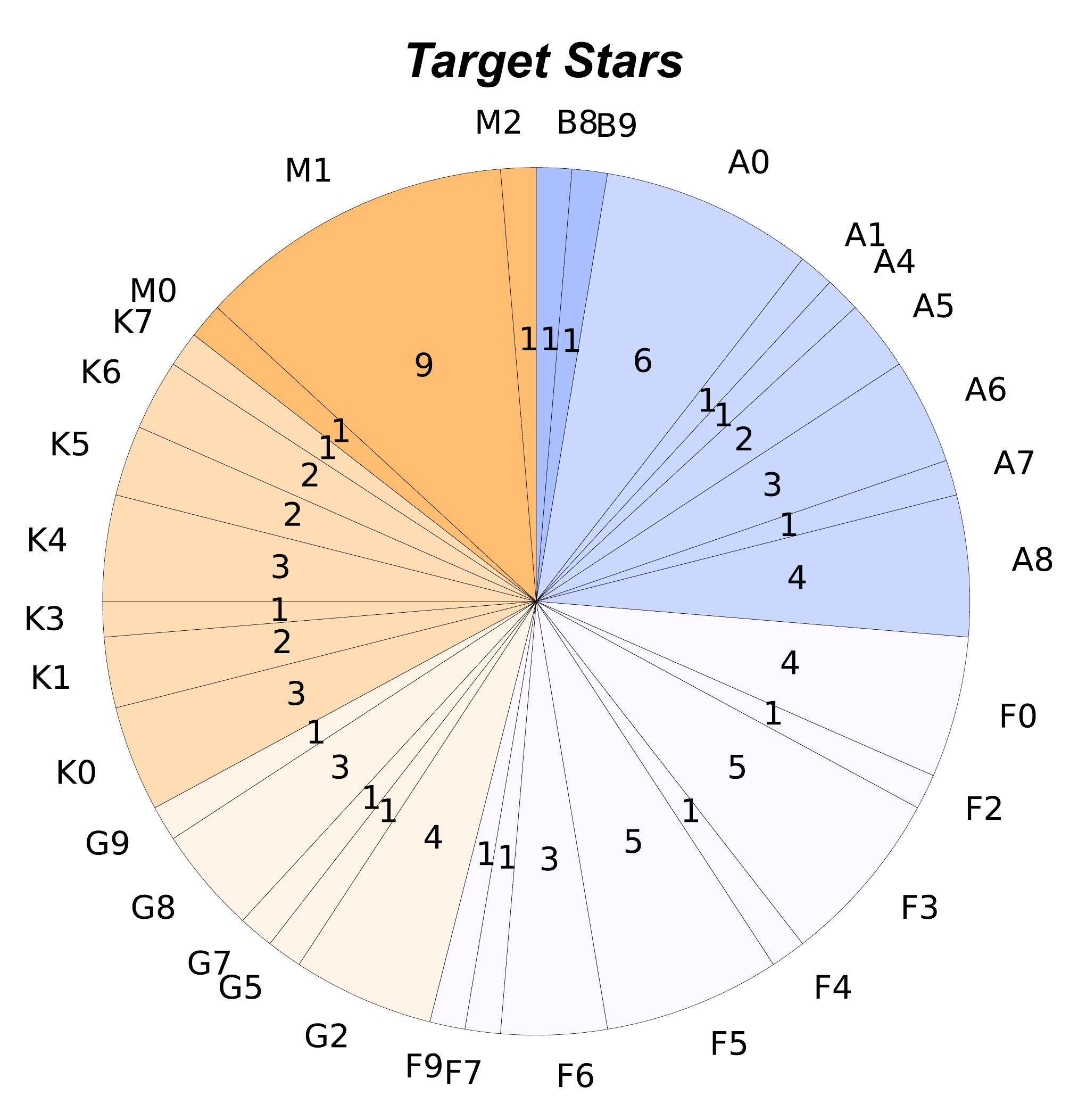}
\end{tabular}
\end{center}
\caption
{ \label{fig:sptypes} Spectral types of the stars observed with either Wedge A0.6 or A1.0 position or both.  {\it Left}: Spectral types of the {\it PSF} stars. {\it Right}: Spectral type of the {\it target} stars. The spectral types are shown outside the pie chart, with  their corresponding number of stars on the wedges. There are $16$ PSF stars and $16$ target stars that are observed with both wedge positions. The color of the wedges roughly reflects the general chromaticity of the class of stars (although chromaticity could vary within a class). Source: the SIMBAD Astronomical Database \cite{simbad}, \url{http://simbad.u-strasbg.fr/simbad}.}
\end{figure}

STIS serves both as a spectrograph and an imaging instrument\cite{woodgate98, grady03}, with several aperture locations capable for coronagraphic imaging (Fig.~\ref{fig:stis-coron}). For our purpose, we obtained all public {\it HST}-STIS coronagraphic imaging observations available in December 2016 from the STScI MAST archive\footnote{\url{http://archive.stsci.edu/hst/search.php}}. A summary of exposure information is listed in Table~\ref{tab:stis-obs}. In order to apply the KLIP algorithm to the archival data, we focused on the most frequently used aperture locations: the Wedge A0.6 and A1.0 positions, and classify the exposures into two categories: targets which are known to have infrared excess in their spectral energy distributions\cite{chen14} (i.e., the ``Targets''), and  references which do not have infrared excess (i.e., the Point Spread Functions, ``PSFs''). The spectral types of the PSF stars and target stars are summarized in Figure.~\ref{fig:sptypes}. For the BAR5 position, we take data the coronagraphic observations of HD 38393 (Proposal ID: 14426\footnote{\url{http://archive.stsci.edu/proposal_search.php?mission=hst&id=14426}}, PI: Debes).

\begin{table}[htb!]
\centering
\caption{Summary of Public HST-STIS Observations as of Dec.~2016.}
\label{tab:stis-obs}
\begin{tabular}{ccc}\hline\hline
   & Proposed Aperture Name & Number of Flat-Fielded Files \\ \hline
1  & BAR10                  & 116             \\
2  & {\bf WEDGE A0.6}             & {\bf 228}            \\
3  & {\bf WEDGE A1.0}             & {\bf 493}             \\
4  & WEDGE A1.8             & 86              \\
5  & WEDGE A2.0             & 39              \\
6  & WEDGE A2.5             & 1               \\
7  & WEDGE A2.8             & 5               \\
8  & WEDGE B1.0             & 37              \\
9  & WEDGE B1.8             & 8               \\
10 & WEDGE B2.0             & 9               \\
11 & WEDGE B2.5             & 116             \\
12 & WEDGE B2.8             & 5\\ \hline
\end{tabular}
\end{table}

\subsection{Alignment of Stellar Images}
The most important step before KLIP subtraction is the alignment of stellar images\cite{choquet14}. The first step is data preparation. We focus on the flat-fielded science files ({\tt *\_flt.fits}), which contains three extensions for each exposure: the {\tt SCI}, {\tt ERR}, and {\tt DQ} extensions, which store the science values, the statistical errors, and data quality flags, respectively\cite{stisDHB11}. We applied a $3\times3$ pixel median filter for pixels having {\tt DQ} flag values of 16, 256, 8192, which correspond to pixels having dark rate $>$ 5$\sigma$ times the median dark level, bad pixel in reference file, and data rejected in input pixel during image combination for cosmic ray rejection, respectively.

We aligned the centers of the stars using Radon Transform-based center determination method described in Ref.~\citenum{pueyo15} after the above-mentioned median filtering. We took the {\tt SCI} extension data, start center searching from the instrumental center given by its {\tt CRPIX1} and {\tt CRPIX2} header values; to make use of the $45^\circ$ and $135^\circ$ major diffraction spikes (Fig.~\ref{fig:raw-images}) for the alignment of the STIS images, we applied an empirical $r^{1/2}$ correction map\cite{stark14} to enhance the pixels of the diffraction spikes at farther separation from the central star. We aligned the centers of stars iteratively using {\it radonCenter}\footnote{We release the code at \url{https://github.com/seawander/centerRadon}.}, which is able to perform Radon Transform with user-selected parameter spaces and regions, therefore greatly boosts the efficiency when comparing with classical Radon Transform. For our STIS data, {\it radonCenter} first searches for the stellar position by performing line integrals along pre-selected azimuthal directions (assuming the greatest value of the line integrals corresponds with the position of the star) for different positions around the initial guess within $\pm 5$ pixels at a precision of 1 pixel, then interpolates and explores the surroundings of the stellar location to reach a precision of 0.01 pixel ($\sim 0.51$ mas). {\it radonCenter} is at least 2 orders of magnitude faster than center searching with classical Radon Transform\cite{pueyo15} when we focus the line integrals only along the $45^\circ$ and $135^\circ$ major diffraction spikes of STIS exposures.

The aligned stellar images are then divided by their exposure times (obtained from the {\tt EXPTIME} header) to have units of counts s$^{-1}$, and cut into sizes of $133\times133$ pixel for the A0.6 exposures,  $213\times213$ pixel for the A1.0 exposures, and $87\times87$ for the BAR5 exposures for further operations.

\subsection{Image Selection}
\begin{figure}[htb!]
\begin{center}
\begin{tabular}{c}
\begin{subfigure}{0.31\textwidth}
\includegraphics[width=\textwidth]{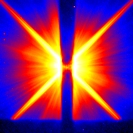}
\caption{Normal Wedge A0.6.}\label{fig2-a}
\end{subfigure}
\begin{subfigure}{0.31\textwidth}
\includegraphics[width=\textwidth]{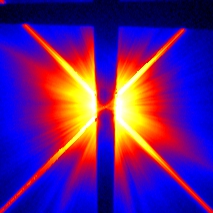}
\caption{Normal Wedge A1.0.}\label{fig2-b}
\end{subfigure}
\begin{subfigure}{0.31\textwidth}
\includegraphics[width=\textwidth]{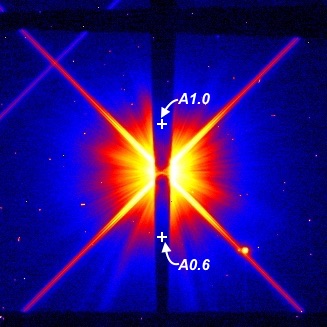}
\caption{Offseted Wedge A0.6/A1.0.}\label{fig2-c}
\end{subfigure}\\
\begin{subfigure}{0.31\textwidth}
\includegraphics[width=\textwidth]{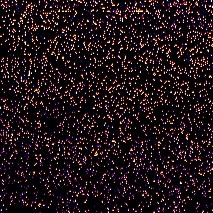}
\caption{Failed.}\label{fig2-d}
\end{subfigure}
\begin{subfigure}{0.31\textwidth}
\includegraphics[width=\textwidth]{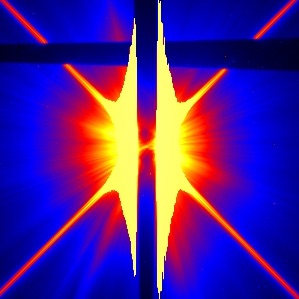}
\caption{Saturated.}\label{fig2-e}
\end{subfigure}
\begin{subfigure}{0.31\textwidth}
\includegraphics[width=\textwidth]{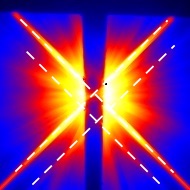}
\caption{Multiple source.}\label{fig2-f}
\end{subfigure}
\end{tabular}
\end{center}
\caption
{ \label{fig:raw-images} Various kinds of STIS exposures in the Wedge A0.6 and Wedge A1.0 archive displayed in log scale with arbitrary units for illustration. The vertical and horizontal dark regions are the coronagraphic wedges of the STIS instrument. The bright lines at the diagonals are the diffraction spikes intrinsic to HST. Only the most commonly used positions, i.e., Panels (a) and (b), are selected to construct our reference and target data cubes. Panel (c) is an offsetted exposure, where the normal positions are marked with white ``+'''s; Panel (d) is a failed exposure, which was caused by the failure of guide star acquisition and no lock is obtained ; Panel (e) is purposely saturating the inner regions to increase the signal-to-noise ratio of possible structures at further separation; Panel (f) contains two sources, which is expected to locate at the intersections of the white dashed lines.}
\end{figure}

For each wedge position (Wedge A0.6, and Wedge A1.0), we carefully inspected the aligned cubes (i.e., the PSF cube and the target cube), then excluded the exposures with too much offset to prevent too much PSF variation (e.g., Fig.~\ref{fig2-c}) -- the total offset from the normal positions, which can be obtained from the sum of the {\tt POSTARG1} and {\tt POSTARG2} headers, should be within 5 pixel\footnote{This is an empirical upper limit we set: beyond 5 pixel the relative movement of the star and wedge will create significant inner working angle change, which causes PSF change consequently.}. The failed exposures (e.g., Fig.~\ref{fig2-d}), exposures with too much saturation (e.g., the vertical bright regions in Fig.~\ref{fig2-e}. The saturated exposures  in principle can be reduced with KLIP after carefully choosing masks which can exclude the saturated pixels, however the shape of the saturated region varies among exposures and this is left for future work), and exposures with multiple sources (e.g., Fig.~\ref{fig2-f}), are manually excluded. We keep only the exposures executed at the most commonly used positions (i.e., Fig.~\ref{fig2-a} and \ref{fig2-b}).

After image selection, we have the exposure information shown in Table~\ref{tab2:exposures}, which summarizes the number of stars and readouts\footnote{In each exposure, there might be several readouts to prevent CCD saturation.} in both positions (Wedge A0.6 and A1.0) and star classifications (PSF and target).

\begin{table}[hbt!]
\centering
\caption{Summary of Exposure Information after Image Selection}
\label{tab2:exposures}
\begin{tabular}{ccc}\hline\hline
Number of Stars (Readouts) & Wedge A0.6 & Wedge A1.0 \\ \hline
PSF                         & 16 (395)   & 30 (820)   \\
Target                      & 20 (1156)  & 68 (1731) \\ \hline
\end{tabular}
\end{table}

\section{Data Reduction}
\begin{figure}[htb!]
\begin{center}
\begin{tabular}{c}
\includegraphics[width=0.45\textwidth]{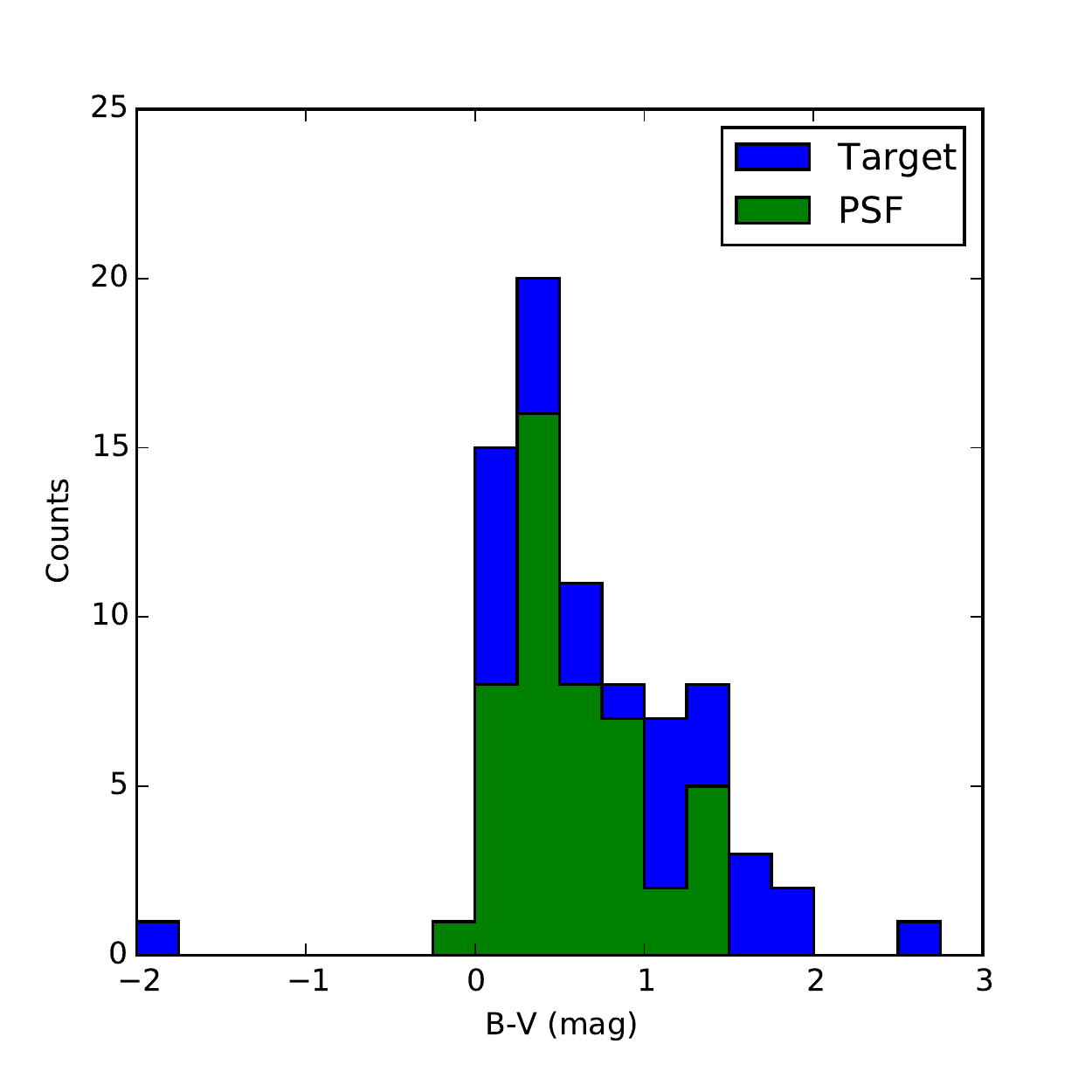}
\end{tabular}
\end{center}
\caption
{ \label{fig:b-v} $B-V$ colors of the STIS Wedge A0.6 and A1.0 combined archive. The blue histogram is for the targets, while the green one is for the PSF stars. Source: the SIMBAD Astronomical Database, \url{http://simbad.u-strasbg.fr/simbad}.}
\end{figure}

Unlike the NICMOS instrument with well-constrained filters as in the ALICE project for archival analysis\cite{choquet14}, the STIS coronagraphic imaging mode has a broad bandpass ($\sim 2000\AA$ -- $\sim10,000\AA$)\cite{stisIHB17}. Therefore in classical RDI reduction with the STIS exposures, reference stars (i.e., PSF stars) are selected based on their $B-V$ colors\cite{grady99, grady00, grady04, roberge05, schneider14}. A summary of the $B-V$ color distribution is shown in Fig.~\ref{fig:b-v}. For our KLIP reduction, we describe our reference selection and KLIP subtraction in this section.

\subsection{Wedge A0.6 and A1.0}
In order to reduce the color-mismatch which creates unrealistic halos (e.g., Fig.~\ref{fig:gj809}), for each target readout, we normalized itself and all the PSF readouts by first applying an algorithmic mask to block the physical wedges and the primary diffraction spikes, then subtracting the mean and dividing the standard deviation of each readout (which normalizes the distribution of the residual pixels, enabling the comparison of different readouts, as well as the calculation of the covariance matrix for the PSF cube which is essential to perform KLIP subtraction), the Euclidean distances were then calculated (although other distances can be calculated, we chose Euclidean distance since KLIP is minimizing variance in a squared distance sense) and the $10\%$ closest matching PSF exposures are used as the corresponding PSF cube of the target readout: we have experimented different levels of closest matchings ($0.1\%, 1\%, 5\%, 10\%, 20\%, 50\%$, and $100\%$) based on the Euclidean distances, and found out that $10\%$ yields the best results with KLIP subtraction with the STIS data.

For each target exposure and readout, we performed the KLIP subtraction by first constructing the KL base using its $10\%$ closest matching PSF readouts, then selecting the first $20$ components, and projected the target exposure to the $20$ components and subtracted the projection from the initial target readout.

For each target, we combined its KLIP results of different readouts and write the final combined result in a FITS file.

\subsection{BAR5 $5\sigma$ Contrast}

\begin{figure}[htb!]
\begin{center}
\begin{tabular}{c}
\raisebox{-0.5\height}{\includegraphics[width=0.42\textwidth]{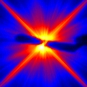}}
\raisebox{-0.5\height}{\includegraphics[width=0.46\textwidth]{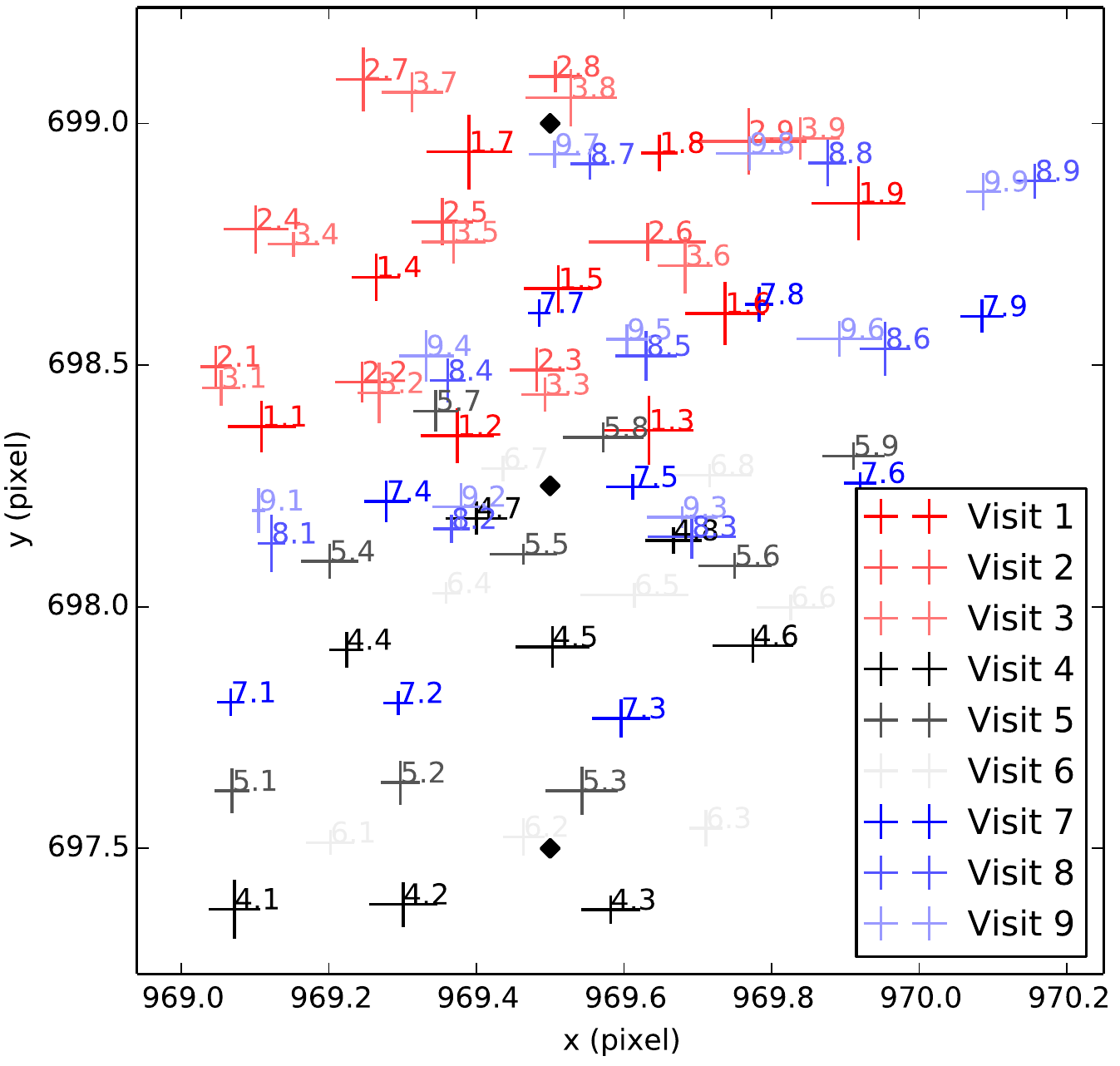}}
\end{tabular}
\end{center}
\caption
{ \label{fig:bar5-raw} {\it Left}: A typical exposure of HD~38393 at the BAR5 position in log scale with arbitrary limits, the central dark nearly horizontal region is the physical mask. {\it Right}: The centers of the 9 BAR5 visits of HD~38393 as determined by {\it radonCenter point}, each ``+'' stands for the mean and standard deviation of the $10$ readouts at each dithering position. The black diamonds group the observations into 3 mutually exclusive sets based on the vertical positions of the readouts.}
\end{figure} 

The original BAR5 exposures of HD~38393 contains a total of 810 of $0.2$-second readouts of HD~38393 at 9 visits with different telescope orientations. A typical BAR5 exposure, and the centers of the visits determined by {\it radonCenter} are shown in Fig.~\ref{fig:bar5-raw}. The 9 visits are executed in three groups in time, each containing 3 different telescope orientations (available from the {\tt ORIENTAT} header): Oct.~2015 ($-129.942^\circ, -105.942^\circ, -90.9426^\circ$), Dec.~2015 ($-68.9435^\circ, -53.944^\circ, -38.9445^\circ$), and Feb.~2016 ($7.05435^\circ, 22.0543^\circ, 37.0564^\circ$). In each visit, the telescope dithers with a $3\times3$ pattern within 1 pixel at separations of $\sim0.25$ pixel (13 mas); at each dithering position, there are $10$ of $0.2$-second readouts.

To calculate the signal-to-noise ratio (SNR) of a planet with known contrast and position observed with the BAR5 location, we generated a point source PSF (hereafter ``synthetic planet'') from the Tiny Tim simulator for STIS coronagraphic imaging\cite{krist11}{\footnote{\url{http://www.stsci.edu/hst/observatory/focus/TinyTim}}}. We added a synthetic planet with known contrast and physical position on the sky to the whole dataset, and performed KLIP subtraction: for the readouts at each telescope orientation (90 images, ``targets''), we treated the readouts at the other $8$ orientations (720 images) as references: we constructed KLIP components with the references, then project the targets onto these components; the projections are then subtracted from the targets. The subtraction results of the 9 telescope orientations are then combined by calculating the median values at each pixel, and SNR of at the location where the synthetic planet is injected then calculated using a $3\times3$ box.

The $5\sigma$ detection limit of the BAR5 occulter is then iteratively calculated by performing a binary search with a synthetic planet at varying contrasts and positions. For each radial separation, its contrast is determined by taking the median of the $5\sigma$ contrasts at $12$ different azimuthal positions ($30^\circ$ azimuthal separation). For detailed calculation of the BAR5 detection limit, refer to the STScI STIS-BAR5 webpage\footnote{\url{http://www.stsci.edu/hst/stis/strategies/pushing/coronagraphy_bar5}}.

\section{Results}
\subsection{Circumstellar Disks}
\begin{figure}[htb!]
\begin{center}
\begin{tabular}{c}
\begin{subfigure}{0.31\textwidth}
\includegraphics[width=\textwidth]{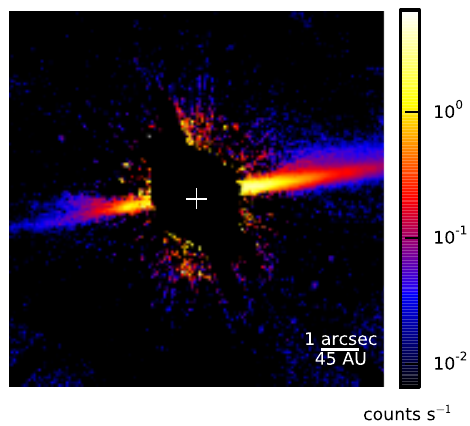}
\caption{HD~15115, Ref.~\citenum{schneider14}.}\label{fig3-a}
\end{subfigure}
\begin{subfigure}{0.31\textwidth}
\includegraphics[width=\textwidth]{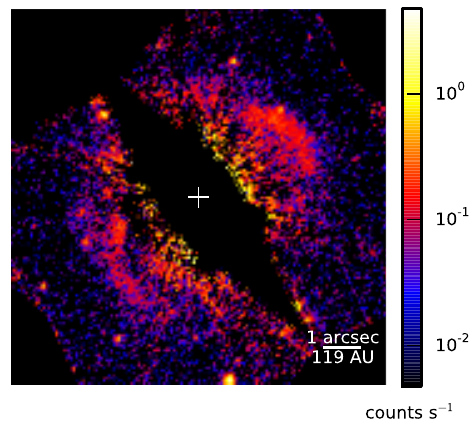}
\caption{HD~163296, Ref.~\citenum{grady00}.}\label{fig3-b}
\end{subfigure}
\begin{subfigure}{0.31\textwidth}
\includegraphics[width=\textwidth]{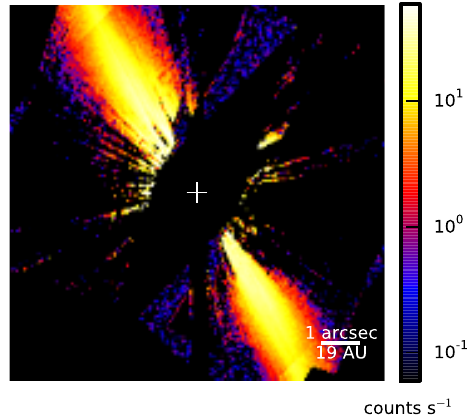}
\caption{$\beta$ Pic, Ref.~\citenum{heap00}.}\label{fig3-c}
\end{subfigure}\\
\begin{subfigure}{0.31\textwidth}
\includegraphics[width=\textwidth]{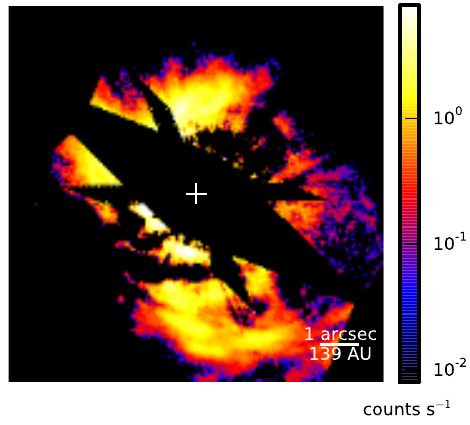}
\caption{AB Aur, Ref.~\citenum{grady99}.}\label{fig3-e}
\end{subfigure}
\begin{subfigure}{0.31\textwidth}
\includegraphics[width=\textwidth]{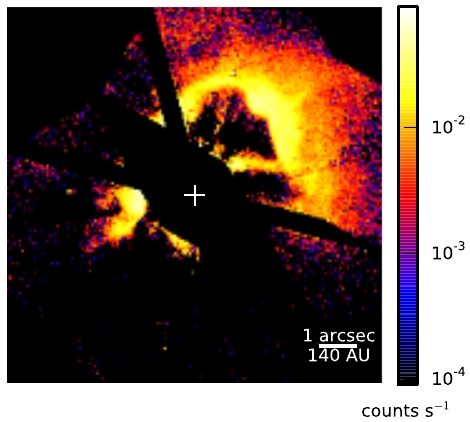}
\caption{DO Tau, Ref.~\citenum{grady04}.}\label{fig3-d}
\end{subfigure}
\begin{subfigure}{0.31\textwidth}
\includegraphics[width=\textwidth]{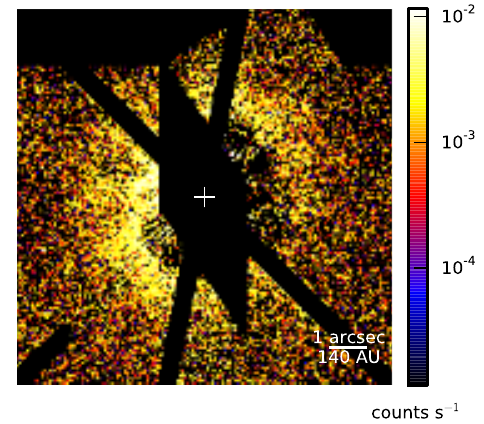}
\caption{DM Tau, Ref.~\citenum{grady04}.}\label{fig3-f}
\end{subfigure}
\end{tabular}
\end{center}
\caption
{ \label{fig:disks} Six circumstellar disk examples in the STIS Wedge A1.0 archive processed with KLIP, the disk morphologies are consistent with the original STIS discoveries. The stellar centers determined by {\it radonCenter} are marked with white ``+'''s.}
\end{figure}

Our KLIP results are consistent with all of the previous attempts both in detection of circumstellar disks ($32$) and null-detection ($41$) . In Fig.~\ref{fig:disks}, we show $6$ of our $32$ KLIP disks (HD~15115, HD~163296, $\beta$~Pic, AB~Aur, DO~Tau, and DM~Tau) that have already been previously imaged with STIS and reduced with classical RDI subtraction\cite{schneider14, grady00, heap00, grady99, grady04}. In addition, we are able to better suppress the PSF at closer-in regions, which is because of the inclusion of an empirical $10\%$ closest matchings. A typical example is the HD~15115 edge-on disk: the faint face-on halo in the first STIS discovery\cite{schneider14} that was reduced with RDI is now removed by KLIP, since KLIP captures the minute difference between PSF wing of the target and its reference(s). 

\begin{figure}[htb!]
\begin{center}
\begin{tabular}{c}
\begin{subfigure}{0.31\textwidth}
\includegraphics[width=\textwidth]{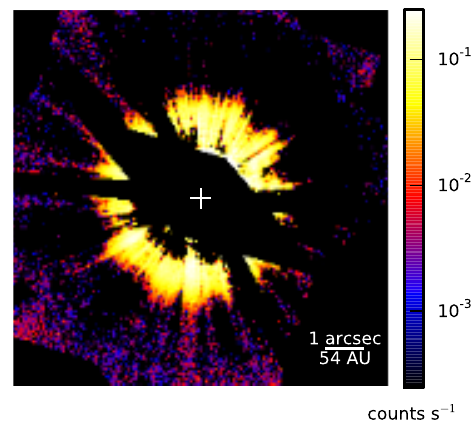}
\caption{TW~Hya, Refs.~\citenum{roberge05} \& \citenum{debes17a}.}\label{fig4-a}
\end{subfigure}
\begin{subfigure}{0.31\textwidth}
\includegraphics[width=\textwidth]{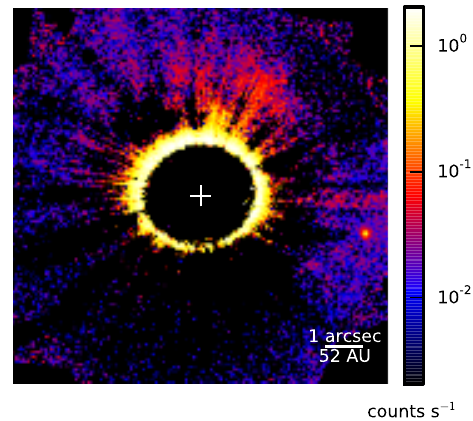}
\caption{HD~181327, Refs.~\citenum{schneider14} \& \citenum{stark14}.}\label{fig4-b}
\end{subfigure}
\begin{subfigure}{0.31\textwidth}
\includegraphics[width=\textwidth]{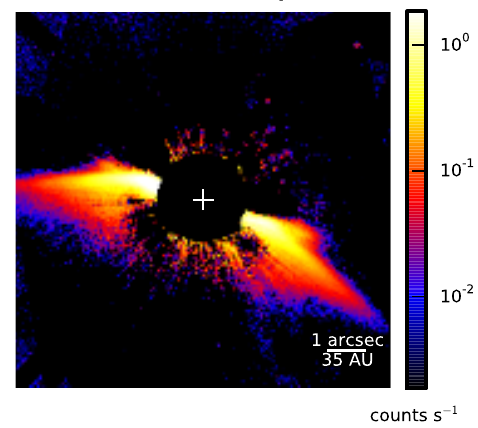}
\caption{HD~61005, Ref.~\citenum{schneider14}.}\label{fig4-c}
\end{subfigure}
\end{tabular}
\end{center}
\caption
{ \label{fig:disks2} Three circumstellar disks examples in the STIS archive processed with KLIP, the disk morphologies are partially changed from their previous discoveries because of the over-fitting as expected.}
\end{figure}

The KLIP results are also recovering some of the disks, although their morphology are modified due to the over-fitting as expected, especially when their morphologies resembles that of stellar PSFs or when the structures are too faint. In these cases, the KLIP projection of the target exposures will overfit the morphology of the disks, changing the morphology of them and forward modeling will have to be implemented. In Fig.~\ref{fig:disks2}, even though the brightest structures remains, the faint structures reported in the original STIS discoveries are not fully recovered because of the mean-subtraction in the data preparation for KLIP.

\begin{figure}[htb!]
\begin{center}
\begin{tabular}{c}
\includegraphics[width=0.45\textwidth]{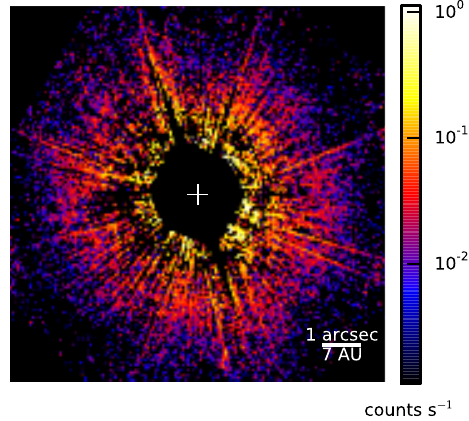}
\end{tabular}
\end{center}
\caption
{ \label{fig:gj809} KLIP subtraction result of GJ~809. The spurious face-on structure is {\bf not} a circumstellar disk because of the color-mismatch. More exposures of the PSF stars matching the color of GJ~809 is needed to better reveal its possible circumstellar disk.}
\end{figure}

In terms of minimizing color-mismatch, KLIP does require more observations of stars with other spectral types. For example, we show in Fig.~\ref{fig:gj809} the KLIP subtraction result of GJ~809. The spurious face-on halo is unlikely to be a disk since its host star being an $M1$ star with $B-V=1.859$, and does not have close match neither in spectral type (Fig.~\ref{fig:sptypes}) nor in $B-V$ color (Fig.~\ref{fig:b-v}).

\subsection{Exoplanet Detection Limit for BAR5}
\begin{figure}[htb!]
\begin{center}
\begin{tabular}{c}
\includegraphics[width=0.5\textwidth]{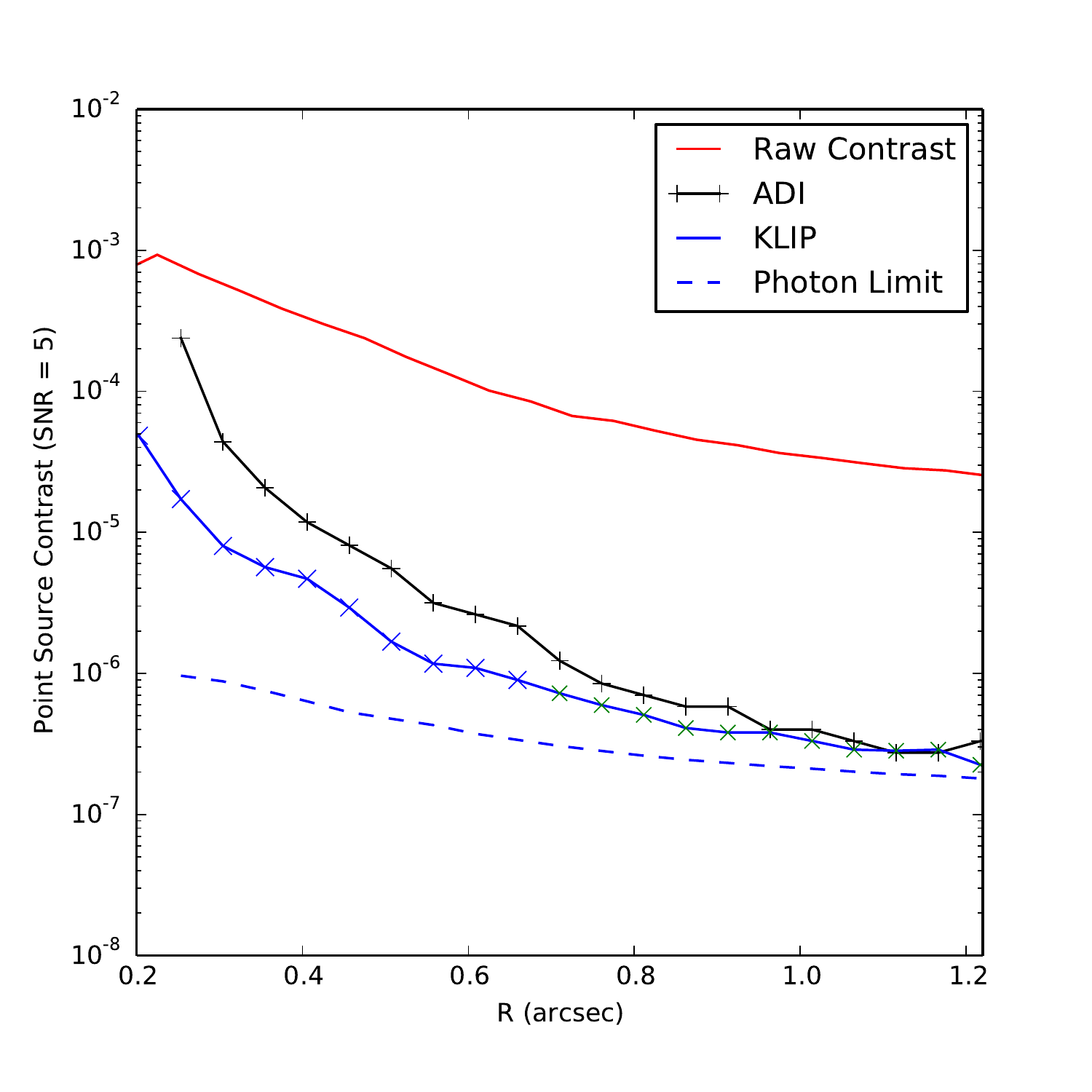}
\end{tabular}
\end{center}
\caption
{ \label{fig:bar5-contrast} $5\sigma$ contrast limits of STIS BAR5. The contrast curve for KLIP is closer to the expected photon limit than the classical ADI subtractions. Exterior of $0.6''$, STIS is able to achieve a contrast better than $10^{-6}$; and STIS is also able to probe an inner working angle of $0.2''$.}
\end{figure}

We performed a binary search of the $5\sigma$ detection limit with both KLIP and classical ADI subtraction with the BAR5 exposures of HD~38393, the contrast limits are shown in Fig.~\ref{fig:bar5-contrast}. 

A reference library of PSFs was generated for each vertical grouping and spacecraft orientation from the other spacecraft orientations to determine the contrast achieved within $1.2''$. We injected synthetic planets to determine the $5\sigma$ contrasts for a range of azimuths. Interior to $0.6''$, for each image (target) in each vertical group containing $270$ readouts, we selected references from the other $269$ readouts which were taken at different spacecraft orientations, and performed KLIP subtraction using the first 170$^\mathrm{th}$ components. Exterior to $0.6''$, we used all images together and selected the first 170 closest matching PSFs from other spacecraft orientations and used these as the final reference library. In this case, all of the 170 components were used for our subtraction. When comparing KLIP with classical ADI subtraction, the KLIP contrast curve is closer to the expected photon limit.

We are able to demonstrate that the STIS BAR5 location is able to image planetary companions at a contrast of better than $10^{-6}$ exterior to $0.6''$. The BAR5 occulter is also able to probe to an inner working angle of $0.2''$ with a contrast better than $10^{-4}$.

\section{Conclusion}
Our post-processing of the STIS observations have confirmed all of the previously reported discoveries (32 circumstellar disks), and obtained the contrast limit of the STIS BAR5 position (better than $10^{-6}$ for separations greater than $0.6''$). With this newly calibrated BAR5 coronagraphic imaging location, STIS is able to image circumstellar disks and exoplanets with an inner working angle of $0.2''$ (Debes \& Ren, in preparation).

For the STIS exposures, previous tedious manual alignment and reference selection is not necessary any more, especially when more advanced post-processing techniques are involved (see Ren et al.~in preparation, for the utilization of non-negative matrix factorization to recover the moprhology of circumstellar disks), and the data cubes we constructed can be used for the reduction of future observations. Our next step for this work will be: 1) performing KLIP reduction on the saturated exposures, 2) quantify our improvement on disk extraction with KLIP over the classical subtraction methods, and 3) apply non-negative matrix factorization of this archive to better recover the morphology of the circumstellar disks.

To achieve better constraints for the STIS circumstellar disks, we have to obtain a more complete coverage of stellar spectral type: on one hand, the number of M-type targets is more than 5 times that of the PSF stars, which significantly limited our goal of selecting the closest PSF matchings before data reduction; on the other hand, the distributions of $B-V$ color for the PSF and target stars do no match, either.

The STIS archive has been and will be used to study the variations in disks (e.g., the yearly variation of circumstellar disk morphology surrounding AU Mic\cite{boccaletti15} and brightness variation of the TW Hya disk\cite{debes17a}). To better study the variations of other disks, especially for the faintest ones, new post-processing methods have to be investigated to both minimize over subtraction and pull out faint disk structure\footnote{We are reducing the whole STIS archive now with non-negative matrix factorization  which is expected to reach these goals (Ren et al.~in preparation).}. With correct morphology of the STIS disks, we will be able to compare them with both archival data (e.g., the NICMOS archival re-discoveries as by Ref.~\citenum{soummer12, choquet14, choquet17}), and current results from ground-based high contrast imaging instruments (e.g., GPI, SPHERE), as well as future telescopes (JWST, WFIRST, etc.).

The combination of observational results at different wavelengths would inform us more about the composition of the disks. Current disk modeling efforts are mainly performed with results from a single instrument (e.g., Refs.~\citenum{hung15, choquet17}), and joint fitting works have been attempted\cite{choquet17}, however there are still degeneracies in grain properties that cannot be easily broken. To break the degeneracies, the simultaneous modeling of multiple wavelengths data should be adopted. In this way, the compositions and therefore origins of the circumstellar disks can be better constrained and explored, which will ultimately contribute to our understanding of the formation and evolution of planetary systems.

\acknowledgments     
Based on observations made with the NASA/ESA Hubble Space Telescope, and obtained from the Hubble Legacy Archive, which is a collaboration between the Space Telescope Science Institute (STScI/NASA), the Space Telescope European Coordinating Facility (ST-ECF/ESA) and the Canadian Astronomy Data Centre (CADC/NRC/CSA). This research has made use of the SIMBAD database, operated at CDS, Strasbourg, France. B.R.~thanks the useful discussions with Christopher Stark, Schuyler Wolff, Jonathan Aguilar, and Alexandra Greenbaum; and computational resources from the support by Colin Norman and the Maryland Advanced Research Computing Center (MARCC) funded by a State of Maryland grant to Johns Hopkins University through the Institute for Data Intensive Engineering and Science (IDIES). E.C.~acknowledges support from NASA through Hubble Fellowship grant HST-HF2-51355 awarded by STScI, operated by AURA, Inc.~under contract NAS5-26555, and support from HST-AR-12652, for research carried out at the Jet Propulsion Laboratory, California Institute of Technology. This research made use of Astropy, a community-developed core Python package for Astronomy\cite{astropy13}.


\bibliography{refs}   
\bibliographystyle{spiebib}   

\end{document}